\documentstyle[aps,preprint]{revtex}
\begin{document}
\preprint{IMSc-96/12/32;~hep-th/9702075}
\title{High Energy Effects on D-Brane and Black Hole Emission Rates}
\author{Saurya Das, Arundhati Dasgupta and Tapobrata Sarkar
\footnote{E-Mail:~saurya,dasgupta,sarkar@imsc.ernet.in}}
\address{The Institute of Mathematical Sciences, \\ CIT Campus,
Chennai (Madras) - 600 113,  India.}
\maketitle
\begin{abstract}

We study the emission of scalar particles from a class of
near-extremal five dimensional black holes and the corresponding
D-brane configuration at high energies. 
We show that the distribution functions and the 
black hole greybody factors are modified in the high energy tail
of the Hawking spectrum 
in such way that the emission rates exactly match. 
We extend the results to charged scalar emission and to four dimensions. 
\end{abstract}
\pacs{04.60.-m,04.62.+v,04.70.-s,04.70.dy,11.25.-w}

\section{Introduction}

Recently, it was shown that microscopic interpretation can be given
to Bekenstein-Hawking entropy of certain stringy black holes. These black 
holes can be identified with elementary or solitonic string states and
the degeneracy of the latter matches with the entropy of the black
hole \cite{sen,vafa,mal}. 
Hawking radiation of scalars has also been understood in
terms of their D-brane description. 
The Planck factor in the Hawking spectrum 
was obtained in \cite{camal,malsu} using the distribution functions of
open string states residing on the D-branes. Shortly, it was shown in
\cite{dasm} that the rate of D-branes decaying into low energy scalars
perfectly matched the Hawking spectrum from the
corresponding black hole. The black holes considered were solutions
of low energy effective action of type IIB string theory
compactified on a five dimensional torus. Their D-brane description
consisted of
$Q_5$ D-5-branes wrapped around $T^5$ and $Q_1$ D-1-branes wrapped
around $S^1$ contained in the $T^5$ and a collection of 
open strings carrying some momentum along $S^1$. 
The situation is equivalent to a single `long' D-1-brane 
wrapped $Q_1Q_5$ times around the $S^1$ \cite{malsu,dm2}. 
The left and right moving massless open string states on this long brane 
constitute two non-interacting one dimensional gases,
approximated by canonical ensembles at low energies. 
A pair of oppositely moving states, each carrying energy $\omega/2$,
can annihilate to form a
closed string state, like the graviton in the internal dimensions,
of energy $\omega$, which cannot reside on the
D-brane and is emitted as a scalar particle. The exact decay rate can be
calculated from the Dirac-Born-Infeld action, which to
leading order is given by :
\begin{equation}
\Gamma_D~=~g_{\mbox{eff}}~\omega~\rho \left(\frac{\omega}{2T_L}\right)
~\rho \left(\frac{\omega}{2T_R}\right)~\frac{d^4k}{(2\pi)^4}~,
\label{db}
\end{equation}
where $\rho \left( \omega/2T_{L,R}\right)~=~1/\left(
\exp(\omega/2T_{L,R}) -1 \right)$ and $g_{\mbox{eff}}$ is related
to the parameters of the
corresponding black hole. $T_L$ and $T_R$ are the effective
temperatures of the left and right moving canonical ensembles. 
In the limit $T_L \gg T_R$, it was shown that \cite{dasm},
\begin{equation}
\Gamma_D~=~A_H~\rho \left( \frac{\omega}{T_H}\right)~\frac{d^4k}{(2\pi)^4}~,
\label{db1}
\end{equation}
where $T_H$ is the Hawking temperature of the black hole \cite{dasm}.
On the other hand, the Hawking spectrum from the black hole is given by
\begin{equation}
\Gamma_H~=~\sigma_{\mbox {abs}} 
~\rho \left( \frac{\omega}{T_H}\right)~\frac{d^4k}{(2\pi)^4}~,
\label{bh}
\end{equation}
where, $\sigma_{\mbox{abs}}$ is the 
greybody factor for the black hole, which for low energy emissions,
is just the area of the event horizon $A_H$.
Substituting, we find that
the rates (\ref{db1}) and (\ref{bh}) match exactly. 

In \cite{malst}, the restriction $T_L \gg T_R$ was
dropped, while still remaining in the near extremal region,
and it was shown in general that for $T_L \sim T_R$, 
\begin{equation}
\sigma_{\mbox {abs}}~=~g_{\mbox {eff}}~\omega~
\frac{e^{\omega/T_H}-1}
{\left(e^{\omega/2T_L}-1\right)~\left(e^{\omega/2T_R}-1 \right)}~,
\label{abs}
\end{equation}
with the Hawking temperature given by
\begin{equation}
\frac{2}{T_H}~=~\frac{1}{T_L} + \frac{1}{T_R}~.
\label{T}
\end{equation}
Once again, it is seen from (\ref{db}) and (\ref{bh}),
that the D-brane and the black hole decay rates match. 

In the above analyses, it was strictly assumed that
the energy of the emitted scalars was vanishingly small. 
Recently, the high energy tail of
the emission spectrum for D-branes as well as black holes was
probed in \cite{kv} although confined to the $T_L \gg T_R$ regime.
The energy $\omega$ was chosen such that $T_R ~,T_H \ll  \omega \ll  T_L$.
In this regime, the right moving open strings were 
treated as a microcanonical ensemble and the corresponding
distribution function was modified to
\begin{equation}
\rho\left(\frac{\omega}{2T_R}\right) \approx \exp \left[ S_R(N_R' -
m) - S_R (N_R') \right]~=\exp(-\Delta S_R)~, 
\label{right}
\end{equation}
where $\Delta S_R$ is the change in the right moving entropy on removal 
of a boson at level $m$ with energy $\omega/2$. 
Here $N'_R$ and $N'_L$ are the left and right moving momenta on the
long D-1-brane respectively (the actual momenta on the 1-brane is
given in terms the quantum number $N_{L,R}=N'_{L,R}/Q_1 Q_5$). 
Now, the black hole entropy is given by \cite{dasm,kv} 
\begin{equation}
S_{BH}~=~2\pi ( \sqrt{N_L'} + \sqrt{N_R'})~.
\label{ent}
\end{equation}
In the limit $T_L \gg T_R$, since
$N_L' \gg N_R'$, we get from (\ref{ent}), $\Delta S_{BH} \approx \Delta
S_R$. Thus, 
\begin{equation}
\rho\left(\frac{\omega}{2T_R}\right) 
\approx \exp \left[ S_{BH}(M- \omega) - S_{BH} (M) \right]~=
~\exp(-\Delta S_{BH}),
\label{rho}
\end{equation}
where $\Delta S_{BH}$ is the change in the entropy of the black hole of
initial mass $M$ after it emits the Hawking particle of energy $\omega$. 

On the black hole side, this change in the distribution function
has been attributed to the back reaction effects which become
important at high energies. In \cite{kraus,kv} this 
was studied by modelling the outgoing particle as a spherical shell
and quantising it. In the WKB approximation, the Hawking factor 
$\rho(\omega/T_H)$ turned out to be precisely the right hand side
of Eq.(\ref{rho}). The left distribution function and the greybody
factor remains unchanged and thus, once again the D-brane and black hole 
emission rates are found to match. 

In this paper, we will relax the condition $T_L \gg T_R$ and
investigate the range $\omega \gg T_{L,R,H}$.
The gas of open strings is treated as a 
microcanonical ensemble in both the left and the right sectors.
We show that the greybody factor gets significantly modified in the
high energy tail of the spectrum. With these, we find that 
the emission rates once again match. Finally, we generalize 
the results for charged scalar emission and to 4-dimensions. 

\section{D-brane Emission Spectrum at High Energies}
\label{emission}

Consider a one-dimensional gas of massless open strings in a box of
length $L$. The total momentum $P$ of the gas is given in terms of
the quantum number $N'$ by $P=2\pi N'/L$ and the energy of a
colliding string by 
\begin{equation}
\omega/2 =2\pi m/L. 
\label{omega}
\end{equation}
For low energy excitations, such that $m \ll \sqrt{N'}$,
the gas is well approximated by a canonical ensemble, and the
distribution function is of the Bose-Einstein form. However, for
higher energies, when 
\begin{equation}
\sqrt{N'} \ll m \ll N', 
\label{micro}
\end{equation}
which amounts to the
excitation energy being much greater than the corresponding
temperature,  the canonical
description is inadequate, and the gas should be described by
a microcanonical ensemble. Since we are interested in the regime
$T_L \sim T_R$ and $\omega$ exceeds these temperatures, the
microcanonical distribution functions should be invoked in the right
as well as the left sectors, which is given by \cite{kv}:
\begin{equation}
\rho \left(\frac{\omega}{2T_{L,R}}\right)~=~
\exp \left[-2\pi \left(\sqrt {N'_{L,R}}- \sqrt{N'_{L,R} -m}~\right)
\right]~.
\label{f}
\end{equation}
From Eqs.(\ref{ent}) and (\ref{omega}), we write
\begin{eqnarray}
\rho \left(\frac{\omega}{2T_L}\right)~
\rho \left(\frac{\omega}{2T_R}\right)~
~&=&~\exp \left[-2\pi \left(
\sqrt{N'_L}+\sqrt{N'_{R}}-\sqrt{N'_L
-L\left(\frac{\omega}{4\pi}\right)}
-\sqrt{N'_{R}-L\left({{\omega }\over
4\pi}\right)}~\right) \right]~ \nonumber \\
~&=&~\exp (-\Delta S_L - \Delta S_R )~=~\exp(-\Delta S_{BH})~.
\label{lr}
\end{eqnarray}
Thus, Eq.(\ref{db}) can be written as
\begin{equation}
\Gamma_D~=~g_{\mbox{eff}}~\omega~ \exp( - \Delta S_{BH})
~\frac{d^4k}{(2\pi)^4}~.
\label{db2}
\end{equation}
In the black hole side, the Hawking factor becomes, on inclusion of
back reaction \cite{kv},
\begin{equation}
\rho\left(\frac{\omega}{T_H}\right) \approx 
\exp \left[ S_{BH}(M- \omega) - S_{BH} (M) \right]~=~\exp(-\Delta S_{BH}),
\label{mod}
\end{equation}
which implies
\begin{equation}
\Gamma_H~=~\sigma_{\mbox {abs}}~\exp(-\Delta S_{BH})~ 
\frac{d^4k}{(2\pi)^4}~.
\label{bh1}
\end{equation}
In the next section, we will calculate $\sigma_{\mbox{abs}}$ and
compare the D-brane and black hole emission rates. 
Note that, in \cite{kv} the left sector did not contribute to
$\Delta S_{BH}$ and the relation between the distribution functions
was 
$$\rho\left(\frac{\omega}{2T_R}\right)~=~
\rho\left(\frac{\omega}{T_H}\right)~.$$
Here, on the other hand, both the sectors become equally important
and contribute to the Hawking factor. 

\section{Black Hole Greybody Factors at High Energies}
\label{grey}

In this section, we calculate the greybody factors for the
$5$-dimensional black hole under consideration for quanta of high
energies. We follow the methods of \cite{unruh,malst,gib}. 
This is appropriate in the energy regime where back
reaction becomes important. 
We solve the Klein-Gordon equation in the background of
the metric given by \cite{malu}:
\begin{equation}
ds^2~=~\frac{1}{\left(f_1f_2f_3\right)^{2/3}}
\left[-dt^2~\left(1-\frac{r_0^2}{r^2}\right)\right]
~+~\left(f_1f_2f_3\right)^{1/3}~
\left[\left(1-\frac{r_0^2}{r^2}\right)^{-1}~dr^2 +
r^2 d\Omega_3^2 \right]~,
\end{equation}
where 
$$
f= \left(1~+~{r_n^2 \over r^2} \right) \left(1~+~{r_1^2
\over r^2} \right) \left( 1~+~{r_5^2 \over r^2} \right)
$$
\begin{equation}
\mbox{and}~~~~h~=~1~-~{r_{0}^{2} \over r^{2}}~.
\end{equation}
The parameters $r_1$, $r_5$ and $r_n$ can be expressed in terms of
the two charges 
$Q_1$, $Q_5$ and the momentum $n$ along the D-1-brane as follows:
$$ r_{1}^{2}={g Q_{1} \over V},~~~ r_{5}^{2}=gQ_{5}, ~~~r_{0}^{2}{\sinh 2
\sigma \over 2}={g^{2}n \over R^{2}V},~~~
r_{n}^{2}=r_{0}^{2}\sinh^{2}\sigma~,$$
where $\sigma$ is a boost parameter. 
The radial part of the 
Klein-Gordon equation for a scalar field $\phi$ corresponding to
the $s$-wave state and propagating in the
background of the above metric is given by
\begin{equation}
{h \over r^{3}}{d \over dr}\left( hr^{3}{dR \over dr}
\right)~+~\omega^{2}fR~=~0~.
\label{kg}
\end{equation}

In our calculations, we relax the low energy condition $\omega
r_5, \omega r_1 \ll 1$, originally
imposed in \cite{malst} and solve the above equation by treating
the new $\omega$ dependent terms that enter due to this relaxation,
as a perturbation over the terms
originally present. The following analysis is valid so long
as $\omega r_1, \omega r_5 <1$, although it need not be
vanishingly small. Towards the end of this section, we show that this
is the relevant range for comparing with the D-brane results. 
Equation (\ref{kg}) is solved by dividing space
into two regions, the near and far zones, and then matching the
solutions at some intermediate region. 
We assume the following relation between the various parameters:
\begin{equation}
r_{0},~r_{n} \ll r_{m} < r_{1},r_{5}~,
\label{res}
\end{equation}
where the near and far solutions are matched at $r=r_m$. 
In the far region, we get from equation (\ref{kg}),
\begin{equation}
{d^{2} \psi \over dr^{2}}~+~\left[
\omega^{2}~+~{-3/4~+\omega^{2}(r_{1}^{2}+r_{5}^{2}) \over r^{2}}
\right] \psi=0~,
\end{equation}
where we have substituted $R=\psi r^{-3/2}$ and the 
restrictions given in (\ref{res}). The term
$\omega^2(r_1^2 + r_5^2)$ was absent in \cite{malst} because of
the low energy condition. 
Defining $\rho = \omega r$, we obtain,
\begin{equation}
{d^{2} \psi \over d\rho^{2}}~-~\left[
-1~+~{3/4~-\omega^{2}(r_{1}^{2}+r_{5}^{2}) \over \rho^{2}}
\right] \psi=0~,
\end{equation}
which has the solution
\begin{equation}
\psi~=~\sqrt{{\pi \over 2} \rho}~\left[ \alpha J_{1-\epsilon}(\rho)
~+~ \beta N_{1-\epsilon}(\rho) \right]~.
\end{equation}
where $\epsilon \equiv \omega^2(r_1^2+r_5^2)/2$.
Now, in the matching region, we use the small $\rho$ expansion for the
Bessel functions, and finally obtain for the solution,
\begin{equation}
R~=~{\sqrt{\pi \over 2}} \omega^{3/2} \left[ {\alpha \over 2} 
\frac {( {\rho \over 2}) ^{- \epsilon}}{\Gamma
(2-\epsilon)}
~+~{\beta \over 2}
\left(\left({ \rho \over 2} \right)^{-\epsilon} \cot \pi (1 - \epsilon)
-\frac{\epsilon}{{\mbox {sin}}\pi(1-\epsilon)}
\left({\rho \over 2}\right)^{-2+\epsilon} 
\right) \right]~.
\label{far}
\end{equation}
On the other hand, the asymptotic expansions of the Bessel
functions yield the solutions
\begin{equation}
J_{1-\epsilon}(\rho)~=~\sqrt{{2 \over \pi \rho}}\cos( \rho-{3\pi \over 4}
+{\pi \epsilon \over 2})~,
\end{equation}
$$N_{1-\epsilon}(\rho)~=~\sqrt{{2 \over \pi \rho}}\sin( \rho-{3\pi \over 4}
+{\pi \epsilon \over 2})~,
$$
which are used to compute the incoming
flux at infinity, given by
\begin{equation}
\Phi_{\mbox{in}}~=~-{\omega \over 4}| \alpha|^{2}~.
\label{fin}
\end{equation}
In this computation, we have dropped a $\beta$ dependent piece. 
From equation ({\ref{far}}), it is clear that the term
multiplying $\beta$ is large for small values of the
perturbation parameter. This implies that   	
$\beta/ \alpha \ll 1$. 

In the near zone, Eq.(\ref{kg}) can be written as
\begin{equation}
{h \over r^{3}}{d \over dr}\left( hr^{3}{dR \over dr}
\right)~+~\omega^{2} \left[ {(r_{n}r_{1}r_{5})^{2} \over r^{6}}
~+~{(r_{1}r_{5})^{2} \over r^{4}}~+~{(r_{1}^{2}+r_{5}^{2}) \over r^{2}}
\right]R~=~0~.
\label{kg1}
\end{equation}
Defining new variables
$v$ and parameters $A,B$ as, 
\begin{equation}
v~=~{r_{0}^{2} \over r^{2}};~~~A~=~ {\omega^{2} \over 4}\left(
{r_{1}r_{5}r_{n} \over r_{0}}\right)^{2};~~B~=~{\omega^{2} \over 4}
\left( {r_{1}r_{5} \over r_{0}} \right)^{2}.
\end{equation}
equation (\ref{kg1}) becomes
\begin{equation}
(1-v){d \over dv} \left( (1-v){dR \over dv} \right)~+~\left[ A+{B \over
v}+{\epsilon  \over {2v^{2}}} \right] R ~=~0~.
\label{shm}
\end{equation}
Notice that close to the horizon, $v \rightarrow  1^-$.
Thus, on writing  
$v~=~1 - \delta$ and expanding the $1/v^2$ term in square
brackets, we obtain the equation for the near region as
\begin{equation}
(1-v){d \over dv} \left( (1-v){dR \over dv} \right)~+~
\left[ A+{B~+ \epsilon/2 \over
v}+ \frac{\epsilon \delta}{2} \right] R ~=~0~,
\end{equation}
Hereafter, we drop the
$\epsilon \delta / 2$ term, which is very small. 
In order to compute the flux of neutral scalars absorbed into the black hole, 
we need to know the near region solution very close to the horizon. In
equation (\ref{shm}), if we make the substitution $y~=~- \ln (1-v)$,
we obtain, in this region, a simple harmonic equation for $R$, namely
\begin{equation}
{d^{2} R \over dy^{2}}~+~(A+B+ \frac{\epsilon}{2})~R~=~0~.
\end{equation}
And the incoming solution is given by
\begin{equation}
R_{\mbox{in}}~=~K~{\mbox {exp}}(-i \sqrt{A+B+\frac{\epsilon}{2}}~\ln (1-v))~.
\label{feq}
\end{equation}
Substituting $z~=~(1-v)$, and writing an ansatz for the solution 
as $R = Kz^{-i(p+q)/2}R_1$, we obtain
\begin{equation}
z(1-z){d^{2}R_1 \over dz^{2}}~+~(1-z)(1-ip-iq){dR_1 \over dz}~+~pqR_1~=~0~.
\end{equation}
This is seen to be a hypergeometric equation in $R$, where we have
defined $p$ and $q$ by the equations
\begin{equation}
(p+q)^{2}~=~4(A+B+ \frac{\epsilon}{2})~~;~~pq~=~B~+~\frac{\epsilon}{2}~.
\label{pq}
\end{equation}
The above equation has the solution 
\begin{equation}
R_1~=~F(-ip,-iq,1-ip-iq,z)~,
\end{equation}
where $F$ is the hypergeometric function, and hence the full solution
for $R$ is given by
\begin{equation}
R~=~Kz^{-i(p+q)/2}F(-ip,-iq,1-ip-iq,z)~.
\label{hyp}
\end{equation}
Next, we express $p$ and $q$ in terms of
the black hole parameters.
Solving the equation (\ref{pq}) yields for $p$ and $q$
\begin{eqnarray}
p~&=&~{\omega r_{1}r_{5} \over 2r_{0}} {\mbox e} ^{\sigma}~+~ 
{\omega r_{0} \over 4}{(r_{1}^{2}+r_{5}^{2}) \over r_{1}r_{5}}{1 \over
{\cosh} \sigma}~, \nonumber \\
q~&=&~{\omega r_{1}r_{5} \over 2r_{0}} {\mbox e} ^{- \sigma}~+~ 
{\omega r_{0} \over 4}{(r_{1}^{2}+r_{5}^{2}) \over r_{1}r_{5}}{1 \over
{\cosh} \sigma}~.
\label{pq'}
\end{eqnarray}
Substituting for $T_{L,R}$, namely
\begin{equation}
T_{L,R}~=~\frac{r_0}{2\pi r_1 r_5}~e^{\pm \sigma}~,
\end{equation}
and using $r_1 \sim r_5$, we get
\begin{equation}
p~=~{\omega \over 4 \pi T_{R}}~+~{\omega r_{0} \over 2 {\cosh}
\sigma}~~{\mbox {and}}~~q~=~{\omega \over 4 \pi T_{L}}~+~{\omega r_{0} \over
{2\cosh} \sigma}~.
\end{equation}
In order to proceed to calculate the absorption cross section, let
us first match the far and near zone solutions at $r=r_m$. 
Extrapolating the near solution given by
equation (\ref{hyp}) to the region of small $v~(\mbox{large}~r)$ yields,
\begin{equation}
R~=~K \left[ {\Gamma(1-ip-iq) \over \Gamma(1-ip) \Gamma(1-iq)}
~+~v_{m}(a~+~b~\ln v_{m}) \right]~,
\end{equation}
where $a$ and $b$ are constants depending on $p$ and $q$. Next,
we expand the right hand side of Eq.(\ref{far}) in
powers of $\epsilon$, and retaining the lowest order terms in
$\epsilon$,
we find the matching condition at $r=r_m$: 
\begin{equation}
{\sqrt {\pi \over 2}} \omega^{3/2} \left[ {\alpha \over 2} \left( 1-
\epsilon~\ln(\rho_m /2) \right)
\right]
~=~K \left[ E+v_m(a+b~\ln v_m) \right]~.
\label{mat}
\end{equation}
where $$E~=~{\Gamma(1-ip-iq) \over \Gamma(1-ip) \Gamma(1-iq)}~ $$
and we have imposed the condition $z \simeq 1$. The matching region is
chosen such that $\omega r_m$ is slightly less than unity. Thus,
the second term on the left hand side of Eq.(\ref{mat}) can be
dropped, and we get the relation as in \cite{malst}
\begin{equation}
\sqrt{\frac{\pi}{2}}~\omega^{3/2}~\frac{\alpha}{2}~=~K~E~.
\end{equation}
Now, let us calculate the absorption cross-section \cite{dasm}. The
flux into the black hole, from Eq.(\ref{feq}) is given by
\begin{equation}
\Phi_{\mbox{abs}}~=~-r_{0}^{2}(p+q) |K|^{2}~.
\label{fabs}
\end{equation}
From Eqs.(\ref{fin}) and (\ref{fabs}), we get
\begin{equation}
\sigma_{{\mbox {abs}}}~=~{4 \pi \over 
\omega^{3}}{\Phi_{{\mbox {abs}}} \over \Phi_{{\mbox {in}}}}
~=~{2 \pi^{2} r_{0}^{2} \over \omega}(p+q){1 \over |E|^{2}}~.
\end{equation}
Using the identity  
$$|\Gamma(1-ix)|^2~=~\frac{\pi x}{\sinh \pi x}~,$$ we get
\begin{equation}
{1 \over |E|^{2}}~=~{2 \pi pq \over p+q}{\exp(2\pi(p+q))~-~1 \over
(\exp(2 \pi p)~-~1)(\exp(2 \pi q)~-~1)}~.
\end{equation}
Now, recalling the expressions for $p$ and $q$, in Eq. (\ref{pq'}), we
see that in the limit when $\omega / T_L,R \gg 1$, we can ignore the 
factors of unity
in the numerator and denominator and finally we are left with the
following expression for the absorption cross-section:
\begin{equation}
\sigma_{\mbox{abs}}~=~\sigma_{\mbox{abs}}^{0}~+~\sigma_{\mbox{abs}}^{1}~,
\end{equation}
where,
$\sigma_{\mbox{abs}}^{0}~=~\pi^3 r_1^2 r_5^2 \omega~$, and the
correction term 
$\sigma_{\mbox{abs}}^{1}~=~4 \pi^{3} \omega r_{0}^{2}r_{1}r_{5}
{\mbox{cosh}}\sigma$.
Thus, we see, following the relation between the various parameters that
we have considered, $$\frac{\sigma_{\mbox{abs}}^{1}}{\sigma_{\mbox{abs}}
^{0}}~\sim \left(\frac{r_0}{r_1}\right)^2~~\ll 1~.$$
Using the definition $g_{\mbox{eff}}~=~\pi^3 r_1^2 r_5^2$, we get,  
\begin{equation}
\sigma_{\mbox{abs}}~=~g_{\mbox{eff}}~\omega~.
\label{gbf}
\end{equation}

Now, let us compare the expressions for the black hole and D- brane
decay rates at the high energy regime that we are considering.
Substituting (\ref{gbf}) in (\ref{bh1}), 
we see that the black hole decay
rate becomes
\begin{equation}
\Gamma_H~=~
~g_{\mbox{eff}}~\omega~\exp \left(- \Delta S_{BH} \right)
{d^4k \over (2 \pi)^4}~
\label{com}
\end{equation}
which is just the D- brane decay rate (\ref{db2}).
It may be noted that this matching cannot be
obtained by naively ignoring the unity factors in 
(\ref{db}),(\ref{bh}) and (\ref{abs}).
This is because of the fact that in the regime of high energy
particle emission that we are interested in, the Planckian distribution
of the Hawking particles is no longer valid and we have
to instead resort to Eqs. (\ref{lr}) and (\ref{mod}). 
Hence our result (\ref{gbf})
effects a subtle match between the black hole and D - brane decay rates at
high energies. It can be shown that in the special case 
$T_L \gg T_R$ ($\sigma \rightarrow \infty$), the
results of \cite{kv} are reproduced. 

A word about the range of validity of the 
above result is in order. As stated earlier,
microcanonical corrections become important when the condition
(\ref{micro}) holds. Using (\ref{omega}) and substituting the expression for
temperature \cite{dasm}, namely
\begin{equation}
T_{L,R}~=~\sqrt{{8E_{L,R} \over L \pi f}}
\end{equation}
in (\ref{micro}), we obtain,
\begin{equation}
{\omega \over T_{L,R}} \gg 1
\label{high1}
\end{equation}
which was the condition under which we had derived (\ref{gbf}).
Hence we see that taking microcanonical corrections into consideration
naturally enforces the high energy condition (\ref{high1}).   
In terms of the black hole parameters, this can be written as 
\begin{equation}
\omega r_5 \gg {r_0 \over r_1}
\label{range1}
\end{equation}
Also, our perturbative analysis is valid so long as $ \omega r_5 < 1$,
which is consistent with the condition $m \ll N_{L,R}'$ of
Eq.(\ref{micro}). 
Hence, the range of $\omega$ for which our calculations are valid is 
\begin{equation}
\frac{r_0}{r_1} \ll \omega r_5 <1~.
\label{range2}
\end{equation}
On the other hand, it is clear that for low energies 
(canonical distribution), $m \ll \sqrt{N'}$, implying 
$\omega r_5 \ll r_0/r_1$. 
Thus, it is sufficient to calculate the greybody factor for
$\omega r_5 \ll 1$, as in \cite{malst}. However, in our case, it becomes
important to look at $\sigma_{\mbox{abs}}$ for higher $\omega$, and
(\ref{range2}) exhausts the range over which the D-brane
distribution functions follow that of microcanonical ensembles. 

\section{Charged Emission Rates Including Back Reaction}

The results of the previous sections can be extended to include charged
scalar emission. 
The decay rate for low energy charged scalar
emission from D-Branes, has been obtained in 
{\cite{gub2}}. The emitted massless graviton
field with polarization along the compact directions now have
a net momentum along the compact $S^1$ direction on which
the 1-brane is wrapped. 
The decay rate is given by,
\begin{equation}
\Gamma_{D}= 
g_{\mbox{eff}}~\frac{(\omega^2-e^2)}{\omega}~\rho\left(\frac{\omega +
e}{2T_L}\right)\rho\left( \frac{\omega -e}{2T_R}\right)
{\frac{d^4 k}{(2\pi)^4}}~.
\label{h}
\end{equation}
Comparing with Eq.(\ref{db}), we find that here 
the energies and the momenta of the left and
right modes are shifted by a factor of $\pm e/2$ respectively.
This ensures that there is a net momentum $e$ in the $S^1$
direction, while the energy of the outgoing particle is
$\omega$. This net momenta along the compact direction gives
rise to a Kaluza-Klein charge $e$ for the space-time scalar. A
mass is also endowed such that $|e|=m$. 
When $ T_{L}>> T_{R}$, and $\omega$ is low, 
the emission rate is,
\begin{equation}
\Gamma_{D}=\frac{ A_{H}(\omega -e)}{\omega}\rho\left(\frac{\omega
- e}{2T_R}\right){\frac{d^4 k}{(2\pi)^4}}~.\label{eq:a} 
\end{equation}
For higher energies, however, the decay rate is modified.
In the regime $T_{L}\sim T_{R}$, $(\omega\pm
e)/2T_{L,R}\gg 1$, the density functions are best approximated
as a microcanonical distribution in this regime. The
expression for the left and right densities are same as that in eq
(\ref{lr}), however with energies $(\omega+e)/2$ and $(\omega-e)/2$ of
the left and right particles respectively. 
The product of the left and right density functions combine to give
\begin{equation}
\rho \left(\frac{\omega + e}{2T_L}\right)~
\rho \left(\frac{\omega -e }{2T_R} \right)~=~
\exp(-\Delta S_{BH})~,
\label{eq:d}
\end{equation}
where now $\Delta S_{BH}$ is given by
$\Delta S_{BH}= S(M,Q)-S(M-\omega,Q-e )$.
Here $M$ is the ADM mass of the black hole and $Q$ it's Kaluza-Klein
charge, proportional to the momentum $N_L-N_R$. Clearly $\Delta S_{BH}$
is the change in
entropy due to the emission of a particle with energy $\omega$ and
charge $e$. Then Eq.(\ref{h}) can be written as
\begin{equation}
\Gamma_{D}= 
g_{\mbox{eff}}~\frac{(\omega^2-e^2)}{\omega}~
\exp(-\Delta S_{BH})~
{\frac{d^4 k}{(2\pi)^4}}~.\label{ch}
\end{equation}

The microcanonical decay rate thus obtained can be reproduced exactly from
field theory, 
following {\cite{kv}}, using the techniques developed in {\cite{kraus}}.
Charged black holes emit charged particles at a rate given by,
\begin{equation}
\Gamma_{H}=~\frac{\sqrt{\omega^2-e^2}}{\omega}~\sigma_{\mbox{abs}}
~\rho\left(\frac{\omega - e}{T_H}\right)~\frac{d^4k}{(2\pi)^4}~.
\label{cr}\end{equation}
The density function is evaluated by computing Bogoliubov coefficients.
These relate the wave function at the horizon to the normal components
of the wave function at $r\rightarrow\infty$. 
Due to the infinite boosts associated with the horizon, the
wavefunction $\phi_h$ is well approximated by the WKB value 
\begin{equation}
\phi_{h}=\exp(\imath S)
\end{equation}
The action $S$ is calculated along the trajectory of the charged shell
which approximates the outgoing charged scalar wave .
The Bogoliubov coefficients are hence,
\begin{equation}
\alpha_{\omega \omega'}=\frac{1}{u(r,e)}\int_{-\infty}^{\infty}e^{\imath\omega t} e^{\imath
S}~dt~~~,~~~\beta_{\omega \omega'}
=\frac{1}{v(r,-e)}\int_{-\infty}^{\infty}e^{-\imath\omega t}e^{\imath S}~dt~,
\end{equation}
where $u(r,e)$ and $v(r,-e)$ give the radial wavefunction of the
positive energy, positively charged particle and negative energy and
negatively charged particle respectively. Since $\omega$ and the action
diverge near the horizon, the saddle point value of the integral
dominates. 
The saddle point value is determined through the following equation, 
\begin{equation}
\frac{\partial S}{ \partial t} \pm \omega =0
\end{equation}
By the Hamilton Jacobi equation, the saddle point
corresponds to ${\cal {H}}^{\mp e}=\mp\omega $, $\cal H$ being the
Hamiltonian of the outgoing particle. 
Since, the trajectory of the charged shell 
is that of a null geodesic in the metric {\cite{kraus}}
\begin{equation}
ds^2=-[N_{t}(M+ {\cal {H}},Q+e)dt]^2 + [dr+ N_{r}(M + {\cal {H}},Q+e)dt]^2, 
\end{equation}
we have,   
\begin{equation}
\dot{r}=N_{t}(M+{\cal H},Q+e) - N_{r}(M+{\cal H},Q+e)~.
\end{equation}
Using this and the fact that ${\partial S/{\partial r}}
= P$ (P is the canonical conjugate momentum), we find;
\begin{equation}
|\alpha_{\omega \omega'}| = \exp( -\mbox{Im}\int_{r_{+0}}^{r_f}P_{+}dr)\;\;\;
|\beta_{ \omega \omega'}| = \exp( -\mbox{Im}\int_{r_{-0}}^{r_f}P_{-}dr)~,
\end{equation}
Where $P_{\pm}$ correspond to the positive energy, positively
charged particle and
negative energy, negatively charged particle trajectories respectively. 
As in the case of {\cite{kv}}, the
positive energy trajectory gives a real value of the
integration, while there is an imaginary contribution from the
other. As $r_0= R(M-\omega, Q-e)-\epsilon$,
\begin{equation}
\mbox{Im}\int_{r_o}^{r_{f}}P_{-}dr = -\pi\int^{-\omega }_{0}{\frac
{ d{\cal H} - \Phi dQ}{\kappa(M+{\cal H},Q-e)}}~=~
-\frac{1}{2} \int_{M,Q}^{M-\omega, Q-e} dS_{BH},
\end{equation}
where we have used $\dot r~dP=d{\cal H} - \Phi dQ$ near the horizon and $d{\cal
H}=(\kappa/2\pi) dS_{BH} + \Phi dQ$, $\Phi$ being the
electromagnetic scalar potential at the horizon. 
Thus
\begin{equation}
|\beta_{\omega \omega'}|^2 = \exp [ S_{BH}(M - \omega,Q - e) 
- S_{BH}(M,Q) ] =
\exp(-\Delta S_{BH}),\;\;\;\; |\alpha_{\omega
\omega'}|=1~.
\end{equation}
The density function, in the high energy approximation is
\cite{kv}, 
\begin{equation}
\rho\left(\frac{\omega - e}{T_H}\right)
\approx{|\beta_{\omega \omega'}|^2 \over |\alpha_{\omega \omega'}|^2}~.
\label{eq:c}
\end{equation}
Therefore, 
\begin{equation}
\rho \left(\frac{\omega - e}{T_H}\right)~=~\exp(-\Delta S_{BH}),
\end{equation}
and from equation ({\ref{eq:d}}), it is seen that this equals the value of 
$\rho (\omega+e/2T_L)~\rho(\omega-e/2T_R)$ obtained from the D-Brane picture.

The greybody factor, $\sigma_{\mbox{abs}}$ for high energies is
determined using the same methods as in the neutral emission
case. The scalar equation is
\begin{equation}
\left(1+\frac{r_1^2}{r^2}\right)\left(1+\frac{r_5^2}{r^2}\right)\left[
\omega^2-e^2 + (\omega\sinh\sigma -
e\cosh\sigma)^2\frac{r_0^2}{r^2}\right]R +
\frac{h}{r^3}\frac{d}{dr}\left(hr^3\frac{dR}{dr}\right) =0
\end{equation}
By making the following transformations as in {\cite{malst}}
\begin{equation}
\omega'^2=\omega^2-e^2,\;\;\;\;e^{\pm\sigma'}=e^{\pm\sigma}\left(\frac{\omega\mp e}{\omega'}\right)\;\;\;\;r_n'=r_0\sinh\sigma',
\end{equation}
we obtain equation ({\ref{kg}}) with $\omega\rightarrow\omega'$ and
$r_n\rightarrow r_n'$. To order $(r_0/r_1)^2$, we again obtain the
$\sigma_{abs}$ as in{\cite{malst}} as
\begin{equation}
\sigma_{abs}=g_{\mbox{eff}}~\omega'~\frac{e^{\frac{\omega -
e\Phi}{T_H}}-1}{\left(e^{\frac{\omega
+e}{2T_L}}-1\right){\left(e^{\frac{\omega
-e}{2T_R}}-1\right)}}
\end{equation}
Here $(\omega + e)/T_L + (\omega - e)/T_{R}=(\omega
-e\Phi)/T_H$ and $\Phi=\tanh\sigma$ is the value of the scalar
potential at the horizon. Thus clearly as $\omega'/T_{L,R}>1$,
$\sigma_{abs}\rightarrow g_{\mbox{eff}}~\omega'$. 
Inserting this value in Eq.(\ref{cr}), it is seen to match with
Eq.(\ref{ch}).  

\section{Scalar Emission in Four Dimensions}

In this section, we briefly comment on the extension of our results to
include high energy emission of scalar particles in the more
realistic case of four dimensions. In
the five dimensional case that we have analysed, inclusion of the high
energy effects did not affect the matching of the black hole and D- brane
decay rates, even when the low energy condition of \cite{malst} was 
relaxed, upto leading order of the perturbation parameter that we considered. 
However, in this case we shall show that the same is not true, and that there is
indeed a leading order correction to $\sigma_{\mbox{abs}}$ which means 
that the decay rates no longer match exactly as one goes beyond the low
energy condition, $\omega r_1 \ll 1$. We shall not
indicate the calculations explicitly, which are essentially in the
same lines as in Section (III), but rather state the main results. 
The results for charged scalar emission
rates may be obtained by a simple extension \cite{gub3}. 
The relevant wave equation whose solution we seek is \cite{gub3,desa}
\begin{equation}
{h \over r^2}{d \over dr}\left( hr^2 {dR \over dr}\right)~+~
\omega^2 fR~=~0~,
\label{ch1}
\end{equation}
where $f~=~\prod_{i=1}^{4} \left(1 + {r_i \over r}\right)$, $r_i$'s
being the parameters of the four dimensional black hole.
This equation can be expanded in powers of $1/r$ in the near and far
regions exactly as we did in the five dimensional case. For the near 
region, keeping the next to leading order term in $1/r$ leads to the
equation
\begin{equation}
{d^2 \psi \over dr^2}~+~\left(1~+~{A \over r}\right)\omega^2
\psi~=~0~,
\label{ch2}
\end{equation}
where   
$$A~=~(r_1~+~r_2~+~r_3) ~~~~~\and~~~~~\psi~=~rR~.$$ 
Interestingly, this equation is formally similar to the Coulomb wave
equation. It has a solution in terms of confluent hypergeometric
equations, namely
\begin{equation}
\psi~=~\alpha F (\eta \rho)~+~\beta G (\eta \rho)~,
\label{ch3}
\end{equation}
Where $\rho~=~\omega r$ and $\eta~=~{-A \omega \over 2}$ is a small
parameter.
The asymptotic form for this expression for large $r$ is given by
\begin{equation}
F~=~g \cos \theta~+~ f \sin \theta~~~~~~~G~=~f \cos \theta~-~g\sin\theta~,
\label{ch4}
\end{equation}
where $f$ and $g$ are constants depending on the black hole parameters,
namely, to first order in $\eta$,  
\begin{eqnarray}
f~&=&~1~+~{\eta \over 2 \rho}  \nonumber \\
g~&=&~- {\eta \over 4 \rho^2}  \\ 
\theta~&=&~\rho~-~\eta \ln 2 \rho - \eta \gamma \nonumber~,
\end{eqnarray}
$\gamma$ being Euler's constant.
The flux of incoming particles calculated from this form for the wave
function at infinity is given by
\begin{equation}
f_{\mbox{in}}~=~{| \alpha |^2 \over 4} \left( -1~-{\eta \over 2 \omega
r}~-~{\eta \over 4 \omega^2 r^2} \right)\left( - \omega~+~{\eta \over r}
\right)~.
\label{ch6}
\end{equation}
It can be seen that most of the terms in the above expression can be
neglected as $r \rightarrow \infty$ and we are left with
\begin{equation}
f_{\mbox{in}}~=~{| \alpha |^2 \over 4} \omega~.
\label{ch71}
\end{equation}
Note that we have dropped a $\beta$ dependent piece in $f_{in}$ as
exactly in the five dimensional calculation it turns out to be extremely
small compared to the $\alpha$ dependent term.

In the near region, the equation (\ref{ch1}) reduces after keeping the
leading order terms in descending powers of ${1 \over r}$ to
\begin{equation}
{h \over r^2}{d \over dr}\left( hr^2 {dR \over dr}\right)~+~
\left( {A \over r^4}~+~{B \over r^3}~+~{C \over r^2} \right) \omega^2
R~=~0~.
\label{ch7}
\end{equation}
In terms of the variables $a$ and $b$ defined according to 
$$\left(a~+~b \right)^2~=~4 \left(A~+B~+C \right);~~~~~~ab~=~B~+C~.$$
very close to the horizon we can once again make the substitution $y~=~-
\log \left( 1~-~v \right)$ so that the equation (\ref{ch7}) reduces to
the simple harmonic equation
\begin{equation}
{d^2 R \over dy^2}~+~\left(a~+~b \right)R~=~0~,
\label{ch8}
\end{equation}
and, just as in the five dimensional case, $a$ and $b$ are related to
the black hole parameters, the relation being given by
\begin{eqnarray}
a~=~{\omega \over 4 \pi T_R}~+~2\pi \omega~T_H \left(r_1
r_3~+~r_2r_3~+~r_1r_2 \right) \nonumber \\
b~=~{\omega \over 4 \pi T_L}~+~2\pi \omega~T_H \left(r_1
r_3~+~r_2r_3~+~r_1r_2 \right)~. 
\label{ch9}
\end{eqnarray}
Again writing the incoming solution as
\begin{equation}
R~=~K \exp \left( -i \sqrt{A~+~B~+~C}~\ln (1~-~v) \right)~,
\label{ch10}
\end{equation}
the incoming flux at the black hole horizon is calculated to be
\begin{equation}
f_{\mbox{abs}}~=~| K |^2 r_0 {(a~+~b) \over 2}~.
\label{ch11}
\end{equation}
Now, matching the near and the far zone solutions as in \cite{malst}
gives the relation
\begin{equation}
{K \over \alpha}~=~{\omega \over E}\left(1~-~{\pi\eta\over 2}\right)~,
\end{equation}
where $E$ is given in terms of gamma functions as before as
\begin{equation}
E~=~{ \Gamma(1-ia-ib) \over \Gamma(1-ia) \Gamma (1-ib)}~,
\end{equation}
so that the calculation for $\sigma_{\mbox{abs}}$ finally yields
\begin{equation}
\sigma_{\mbox{abs}}~=~{4 \pi \over \omega^2}{f_{\mbox{abs}} \over 
f_{\mbox{in}}}~=~ 
\sigma^0_{\mbox{abs}} \left[ 1~-~\pi \eta \right]~.
\label{ch12}
\end{equation}
where $\sigma_{\mbox{abs}}^0~=~4 \pi^2 r_1r_2r_3 \omega$ at high
energies. 
Hence we see from (\ref{ch12}) that there is a order $\eta$
correction to $\sigma_{\mbox{abs}}$, in contrast to the five dimensional
case where this correction was negligible. 
This, along with the
microcanonical condition implies that the D-brane and black hole
decay rates match only in the energy regime given by 
$$ \sqrt{\frac{r_0}{r_1}} \ll \omega r_1 \ll 1~,$$
where the emission rates in four dimensions is given by
$$\Gamma_H~=~\Gamma_D~=~4 \pi^2 r_1r_2r_3 
~\omega~\exp(-\Delta
S_{BH})~\frac{d^3k}{(2\pi)^3}~.$$

The difference between the five and four dimensional
cases that we have dealt with is also apparent from the 
general analysis of \cite{desa}. The far zone equation that was 
effectively the source of the difference in the two different dimensions
can be written to leading order, for $D$ dimensions as, 
\begin{equation}
{d^2 \psi \over d \rho^2}~+~\left[ 1~-~\frac{(D-2)(D-4)}{\rho^2} \right]
\psi~=~0~,
\label{alvis1}
\end{equation}
where $\rho~=~\omega r$, and $R~=~r^{-{D-2 \over 2}}\psi$. 
The general solution of the above equation is
\begin{equation}
F~=~\sqrt{{\pi \over 2}}\rho^{{1 \over 2}}J_{(D-3)/2}( \rho)~.
\label{alvis2}
\end{equation}
In five dimensions, the addition of an interaction terms also of
the form $1/\rho^2$) simply modifies the order of the Bessel
function. The resulting corrections in $\sigma_{\mbox{abs}}$ is
negligible. However, in four dimensions, there is a new
$1/\rho$ term, which gives rise to the Coulomb wave function and a
leading order correction in the final result. 

\section{Discussions}

In this paper, we have compared black hole and D-brane decay rates
for neutral and charged scalar emission at high energies. The
microcanonical picture was used in the D-brane side, which naturally
incorporates the condition $\omega/T_{L,R} \gg 1$. This condition
was crucially used in calculating the black hole greybody factor.
In five dimensions, the decay rates match for all values
of energy consistent with the microcanonical picture, while in four
dimensions, they match in a restricted range. It would be
interesting to investigate this difference and see if a more careful
analysis can bring about an exact matching in four dimensions. 

In our calculation of high energy greybody factors, 
it was assumed that there were no explicit
back reaction effects as in the case of the Hawking spectrum. This
can be justified as follows: the modified black hole metric due to
back reaction of the shell can be approximated by the original
metric with a shift in the ADM mass of the black hole by $\omega$.
The corresponding effect on the D-brane is the reduction of the
excitation energy of the gas of open strings i.e. $ E = E_L +E_R
\rightarrow E- \omega$. Since $E$ can be written as
$$E~=~\frac{\pi~r_0^2~\cosh 2\sigma}{8G_5}~,$$
the parameters $r_0$ and $\sigma$ are changed accordingly. It can
be shown that these changes result in a correction also of order
$(r_0/r_1)^2$ in $\sigma_{\mbox{abs}}$ and hence can be ignored. 

\begin{center}
{\bf ACKNOWLEDGEMENTS}
\end{center}

We thank S. R. Das, T. Jayaraman, P. Majumdar, 
P. Ramadevi and G. Sengupta for discussions. One of us (A.D) would
like to thank G. Date for discussions. We thank the organisers of
the 1996 Puri Workshop, for hospitality, where part of the work was
done.

\end{document}